Comment on
" Similarity analysis in magnetohydrodynamics:effects of Hall and ion-slip currents on free convection flow and mass transfer of a gas past a semi-infinite vertical plate", A.A. Megahed, S.R. Komy, A.A. Afify [Acta Mechanica 151, 185-194 (2001)]

Asterios Pantokratoras
Associate Professor of Fluid Mechanics
School of Engineering, Democritus University of Thrace,
67100 Xanthi – Greece
e-mail:apantokr@civil.duth.gr

In the above paper is investigated the boundary layer flow of an electrically conducting fluid over a vertical, stationary plate placed in a calm fluid. The effects of Hall and ion-slip currents are taken into account. The boundary layer equations are transformed into ordinary ones using a scaling group of transformations and subsequently are solved numerically. However, there are two fundamental errors in the above paper which are presented below:

The boundary conditions for this flow are (equations 13.1 and 13.2 in the above paper)

at y = 0:  $u=v=0$ , $T=T_w$ , $C=C_w$          (1)

as $y \rightarrow \infty$  $u=w=0$, $T=T_\infty$ , $C=C_\infty$          (2)

where u, v and w are the velocity components in the x, y and z directions, T is the fluid temperature and C is the concentration of a substance. The initial dimensional boundary layer equations are nondimensionalized using the following non-dimensional variables

$$\bar{x} = \frac{xU_\infty}{\nu}$$          (3)

$$\bar{y} = \frac{yU_\infty}{\nu}$$          (4)

$$\bar{u} = \frac{u}{U_\infty} \tag{5}$$

$$\bar{v} = \frac{v}{U_\infty} \tag{6}$$

$$\bar{w} = \frac{w}{U_\infty} \tag{7}$$

$$\theta = \frac{T - T_\infty}{T_w - T_\infty} \tag{8}$$

$$\varphi = \frac{C - C_\infty}{C_w - C_\infty} \tag{9}$$

where ν is the fluid kinematic viscosity. In the transformed equations two Grashof numbers appear and are defined as follows in the list of symbols

$$Gr = \frac{g\beta(T_w - T_\infty)\nu}{U_\infty^{\ 3}} \tag{10}$$

$$Gc = \frac{g\beta^*(C_w - C_\infty)\nu}{U_\infty^{\ 3}} \tag{11}$$

where g is the gravitational acceleration, β is the fluid thermal expansion coefficient and $\beta^*$ is the fluid concentration expansion coefficient. In the list of symbols the velocity $U_\infty$ has been defined as the free stream velocity. However, in the present problem there is no free stream velocity because the ambient fluid is stagnant (see boundary conditions in equation 2). Is it possible to non-dimensionalize a problem with a non-existent quantity?

It should be noted that the problem of free convection along a vertical isothermal plate, placed in a calm fluid, is a classical problem in fluid mechanics and has been solved by Ostrach (1953) for Pr number from 0.1 to 1000. For this problem the Grashof number is defined as follows (Jaluria, 1980, page 24, Bejan, 1995, page 166, Schlichting and Gersten, 2003, page 91)

$$Gr = \frac{g\beta(T_w - T_\infty)x^3}{\nu^2} \tag{12}$$

We see that the Grashof number given by equation (12) is completely different from that given in equation (10).

It is known in boundary layer theory that velocity and temperature profiles approach the ambient fluid conditions asymptotically and do not intersect the line which represents the boundary conditions. Asymptotically means that the velocity and temperature gradient at large distance from the plate is zero. This demand exists also in the above work taking into account the boundary conditions (40.2). However, this does not happen in the above paper. There are in total 18 velocity profiles, all intersect the horizontal axis with a steep angle at $\eta=3$ and therefore are wrong. These velocity profiles are similar to those developed in the flow between two vertical parallel plates and not to flow along a single plate. Some velocity and temperature profiles that approach the ambient conditions correctly (asymptotically) in a boundary layer flow are shown in Arpaci and Larsen (1984, page 154), in Cebeci and Bradshaw (1988, page 42), in Kakac and Yener (1995, page 47), in Bejan (1995, page 43), in Incropera and DeWitt (1996, page 290), in Oosthuizen and Naylor (1999, page 62), in Schlichting and Gersten (2003, pages 215, 265 and 281) and in White (2006, page 80). It is clear that the profiles which do not approach the horizontal axis asymptotically and intersect it, are truncated due to a small calculation domain used. The authors used for all cases a calculation domain with $\eta_{max}=3$. However this calculation domain was not sufficient to capture the real shape of profiles and a wider calculation domain, greater than 3, should be used. This is the second error in the above paper.

REFERENCES

1. Arpaci, V. and Larsen, P. (1984). Convection Heat Transfer, Prentice-Hall, New Jersey.

2. Bejan A., (1995). Convection Heat Transfer, $2^{nd}$ ed., John Wiley & Sons, New York.

3. Cebeci, T. and Bradshaw, P. (1988). Physical and Computational Aspects of Convective Heat Transfer, Springer Verlag, New York.

4. Incropera, F. and DeWitt D. (1996). Fundamentals of Heat and Mass Transfer, $4^{th}$ edition, John Wiley & Sons, New York.

5. Jaluria, Y. (1980). Natural Convection Heat and Mass Transfer, Pergamon Press, Oxford.

6. Kakac, S. and Yener, Y. (1995). Convective Heat Transfer, $2^{nd}$ ed., CRC Press, Boca Raton.

7. Megahed, A.A., Komy, S.R. and Afify, A.A. (2001). Similarity analysis in magnetohydrodynamics:effects of Hall and ion-slip currents on free convection flow and mass transfer of a gas past a semi-infinite vertical plate, Acta Mechanica, Vol. 151, pp. 185-194.

8. Oosthuizen, P. and Naylor, D. (1999). Introduction to Convective Heat Transfer Analysis, McGraw-Hill, New York.

9. Ostrach, S. (1953), An analysis of laminar free convection flow and heat transfer about a flat plate parallel to the direction of the generating body force. NASA Tech.. Rep. 1111.

10. Schlichting, H. and Gersten, K. (2003). Boundary layer theory, $9^{th}$ ed., Springer, Berlin.

11. White, F. (2006), Viscous Fluid Flow, $3^{rd}$ edition, McGraw-Hill, New York.